\documentstyle[epsfig,longtable]{aipproc}

\begin{document}
\title{Distribution of binary mergers around galaxies}

\author{Tomasz Bulik and  Krzysztof Belczy{\'n}ski}
\address{
Nicolaus Copernicus Astronomical Center,
Bartycka 18, 00-716 Warszawa,Poland}

\maketitle

\begin{abstract} 
We use a  stellar binary population synthesis
code to find the lifetimes and velocities of  several types of
possible GRB progenitors: double neutron stars, black hole
neutron stars, black hole white dwarfs, helium star mergers. 
Assuming that they are born in different types of galaxies we
compute their spatial distribution and compare it  with the
observed locations of GRB afterglows within  their hosts. We 
discuss constraints on the compact object merger model of GRBs
imposed by this comparison and find that the observations of
afterglows and their host galaxies appear inconsistent with the
GRB compact object merger model.
 \end{abstract}

\section*{Introduction}

In the last few years the astronomical community has moved much
closer to unveiling the nature  of GRBs. The discovery of
afterglows and identification of host  galaxies for several
bursts clearly links GRBs to  some type of 
 stellar events. Yet the nature of
these events is unknown, and we still do not know what the GRB
central engines are. Observations of GRB host galaxies and
precise locations of  GRBs within hosts provide a tool to test
some of the possible central engine models. In this paper we
discuss the consistency between the current observations and the
results of binary  population synthesis.

\section*{The Model}

The population synthesis code used here is described in
\cite{bbrhere}. One of the most important parameter
deterimining the properties of the populations of binaries is
the kick velocity a newly formed compact object receives at 
birth. However, several studies  \cite{cordeschernoff,fryer98} 
indicate that the distribution of kick velocities  consists of
two components: a low velocity with the width of approximately
$200\,$kms$^{-1}$, and a high velocity with the characteristic
velocity around $800\,$kms$^{-1}$.  About 80\% of the kicks are
drawn from the first  component.  It is also known that the
production rate  of compact object binaries falls off
exponentially with increasing kick velocity, see Fig.~1 in
\cite{bbrhere}. Thus the population  of compact object binaries
will be dominated by the objects formed in the systems that
received the kicks  drawn from the low velocity component of the
distribution. In the following we will consider the properties
of  the compact object binaries for the case when the kick
velocity is drawn from a Gaussian distribution with the width 
$200\,$kms$^{-1}$.

Little is known a priori about the masses and gravtational
potentials of host galaxies where GRB progenitrs reside.
Therefore to find the expected distribtion of merger  sites
around galaxies  we consider two extreme cases: propogation in a
potential of a large galaxy like the Milky Way and propagation
in the empty space~\cite{bbzmn}.

\begin{figure}[t!] 
\centerline{\epsfig{file=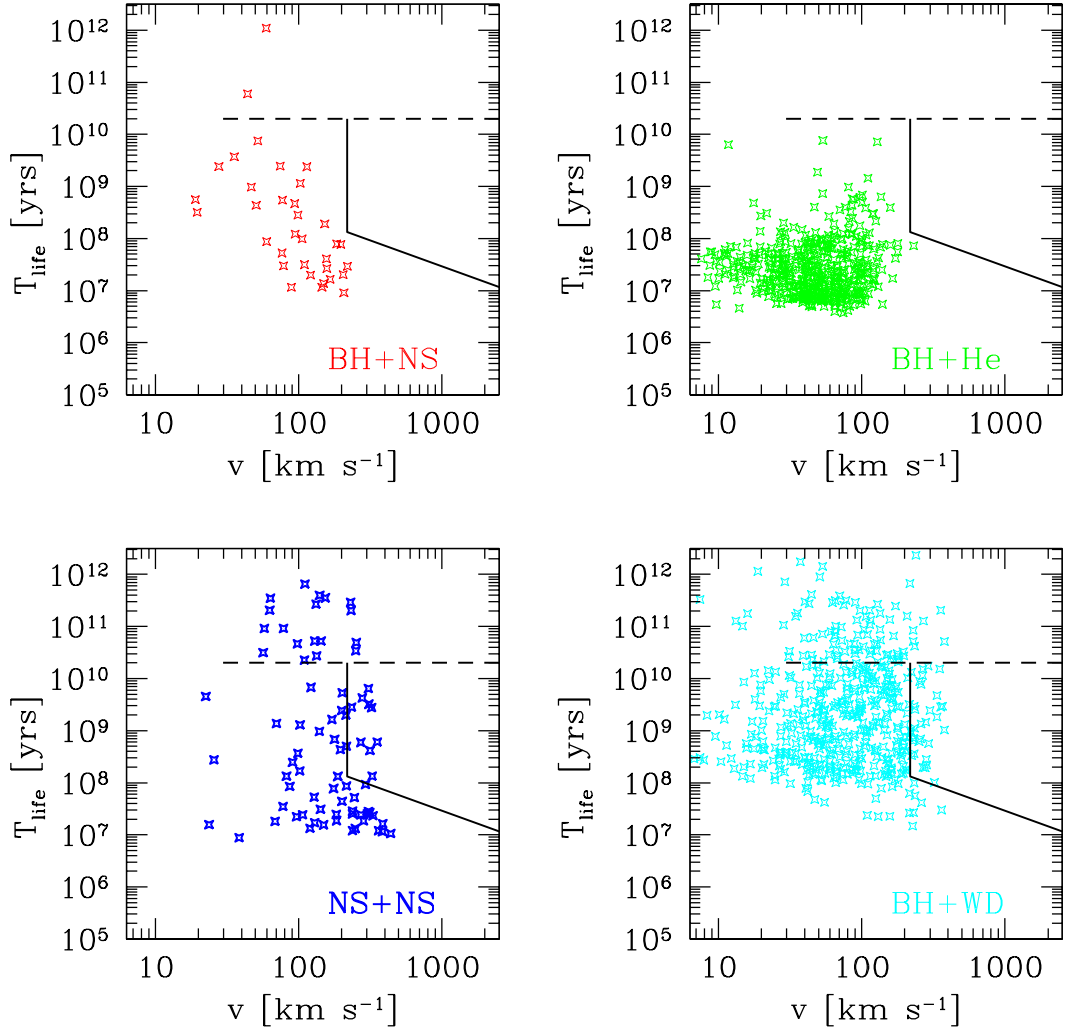,width=\textwidth}}
\vspace{10pt}
\caption{
The distribution of different type of compact object binary
mergers in the space spanned by the binary center of mass velocity and 
the binary lifetime (from ZAMS to final merger). 
The horizontal dashed line corresponds to the Hubble time (15Myrs). 
In the region for $t_{merge} < 15\,$Myrs we present two solid lines: 
the vertical corresponding to $v=200\,$km~s$^{-1}$ -- approximately the 
escape velocity from a galaxy, and the line corresponding to a constant
value of $v\times t_{merge} =30\,$kpc. Together these lines define the
region in the parameter space with systems that can escape from the host 
large galaxy.}
\label{fig1}
\end{figure}

\begin{figure}[t!] 
\centerline{\epsfig{file=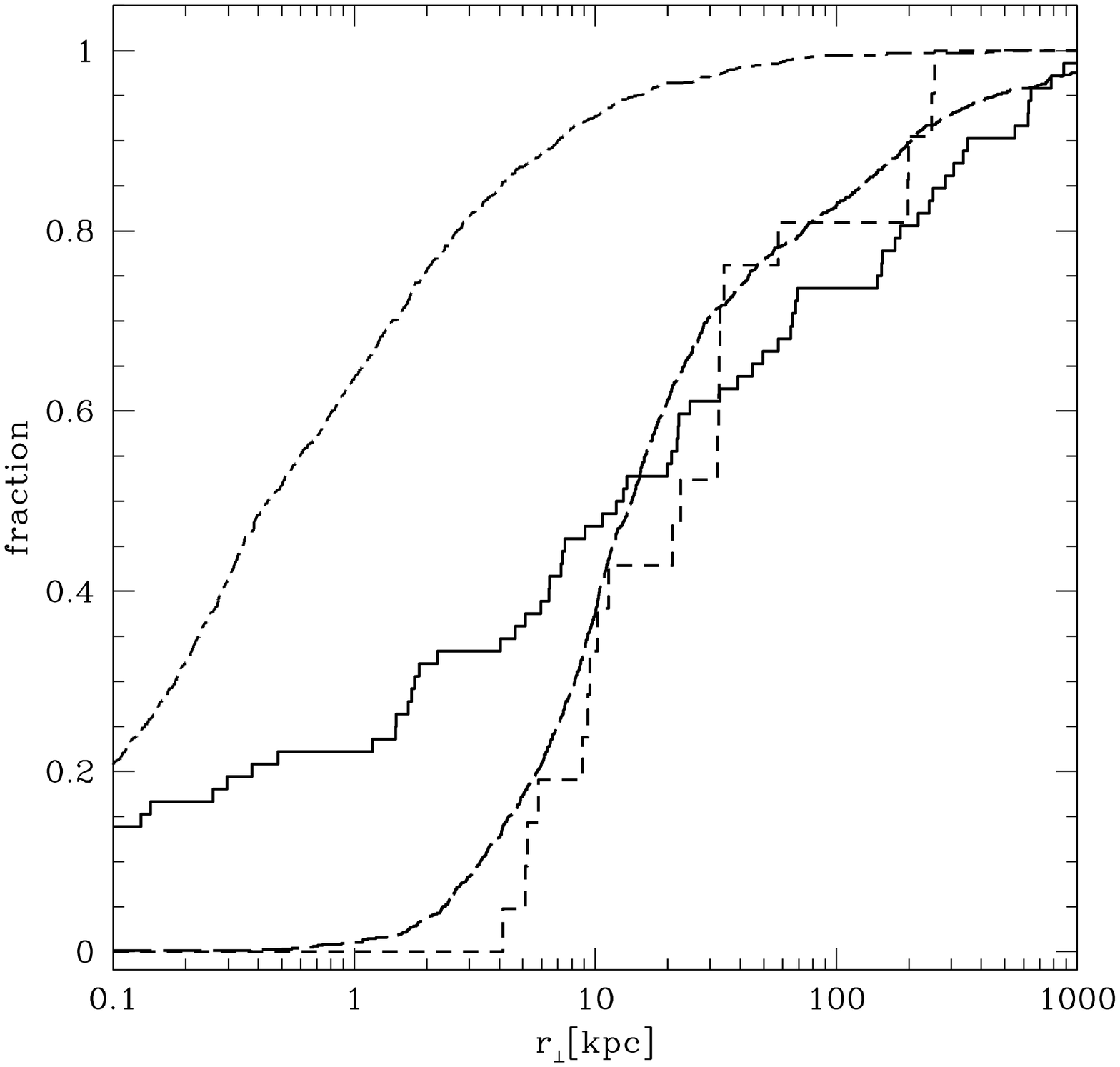,width=0.48\textwidth}
\epsfig{file=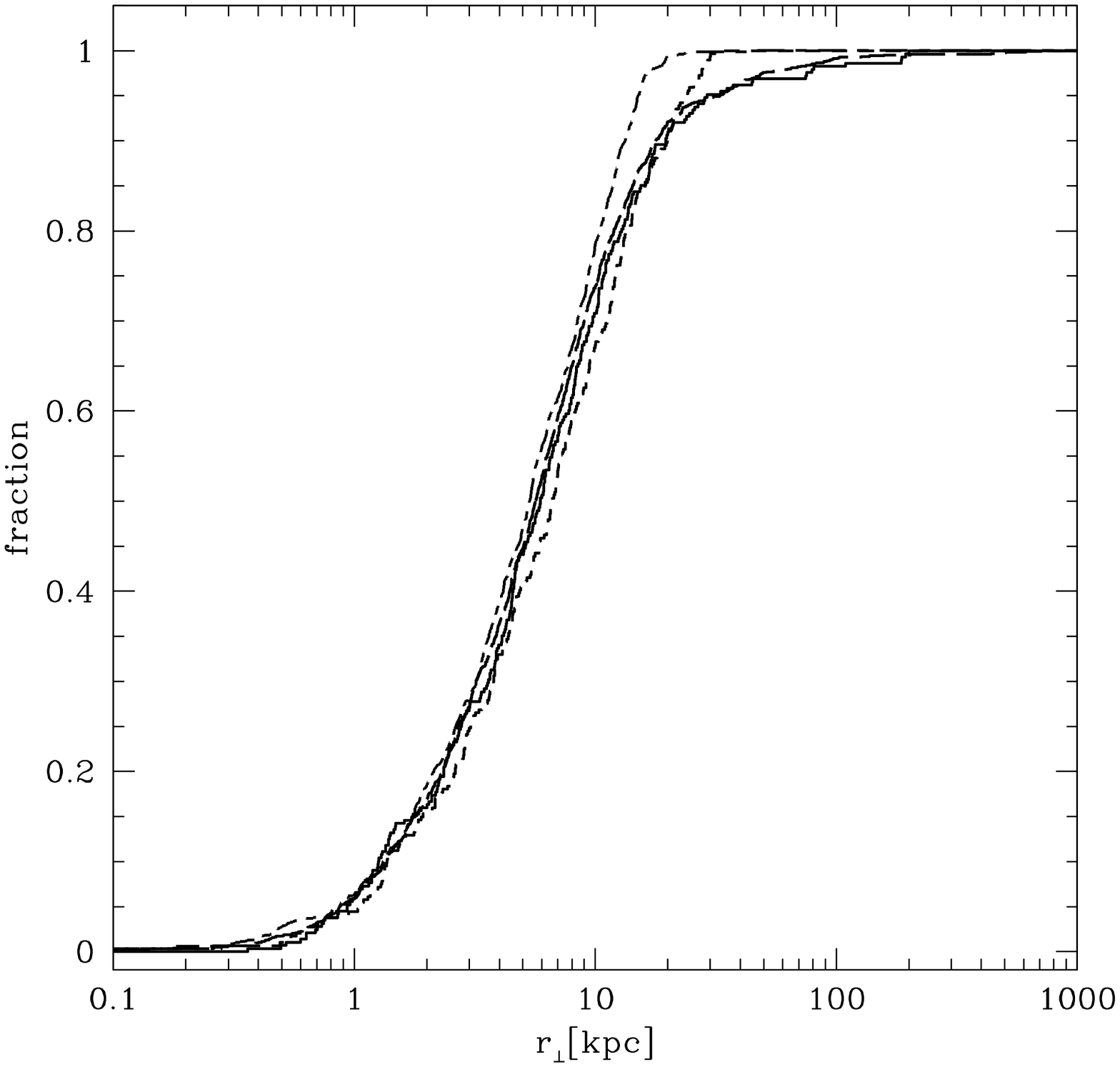,width=0.48\textwidth}}
\vspace{10pt}
\caption{Distributions 
of different types of compact object binaries around
their hosts. Left panel shows the case of prpagation in empty space
and the right panel shows the  case of propagation in a potential of  
a massive galaxy.
The solid, short-dashed, long-dashed and dashed-dotted lines correspond 
to NS-NS, BH-NS, WD-BH, and He-BH mergers respectively. 
}
\label{fig2}
\end{figure}

\section*{Results}

In Figure~1 we present the distribution of center of mass velocities 
gained by systems in the supernova explosions in the galactic 
potantial and binary lifetimes (the time binary takes to evolve 
from ZAMS to final merger of two components) for four types of 
compact object binaries. 

For the case of propagation in a potential of a massive galaxy  only 
a fraction of NS-NS binaries will be able to escape from their
host galaxies. The BH-NS binaries tend to stay in the galaxy.
Here, we have assumed that the  kick velocity does not depend on
the nature of the compact object  formed in the supernova.
However, it has been argued that the kicks back holes receive
should be smaller than those of the neutron stars. This is
discussed in more detail in \cite{bbz}. The helium star mergers
stay in the host galaxies, while some of white dwarf black
hole mergers have a chance of  taking place outside the host,
provided that the escape velocity from a given galaxy is not too 
large.

In the case of propagation in empty space quite a large number of 
binaries of any type will be able to escpae from their birthplace.

In Figure~2 we present the cumulative distributions of the distances
(projected on the sky) between mergers sites and the host galaxies.
In the case of propagation in empty space (left panel  of Figure~2)
most of the mergers take place far outside the host. 
In the case of propagation in the potential of a massive galaxy
(right panel of Figure~2) black hole neutron star mergers and helium star
mergers take place inside the host, and only a small, but not negligible
fraction of double neutron stars and black hole white dwarf mergers 
happen outside the host.

\section*{Discussion}

We have learned at this conference \cite{Fruchter} that  the
afterglow locations coincide  with galaxies, and that 
typically there are intense  star forming processes in these
galaxies. 

In the compact object merger model of GRBs one has to take into
account the fact there are significant delays between  the
stellar formation and the time of merger. This delays consists of
the stellar evolutionary time leading to supernavae  explosions
and formation of the compact onbject binary and the evolution of
the compact object binary due to gravitational wave energy loss.
The distribution of the delay times is rather wide and varies for 
different types of binaries \cite{bbrhere}, see Figure~1.  
In the case of helium star mergers these delays can be as short as
a few million years. 

Assuming that GRBs are related with
NS-NS, BH-NS, or BH-WD mergers we do not expect any
correlation between the GRB sites and star formation because the
star formation  processes could have ceased by the time the
merger happens. 
Thus, within this model GRB rate should be
proportional to the luminous mass in the Universe. As most of the
luminous mass is concentrated in massive galaxies we expect to
find GRBs within such galaxies. However, this should be  typical
galaxies and we tend to find GRB hosts in small, star forming
ones \cite{Fruchter}.

Let us now consider  the case when the delays between the stellar
formation and the merger events are shorter than the star
forming episode itself. Naturally, a correlation between  the
GRB sites and star forming galaxies exists. However, the
observed  host  galaxies in this case are typically small. Thus,
as shown above, a significant fraction of the mergers should take
place  outside the host galaxies and we should be finding GRBs
with no underlying host galaxies. 

One can also argue that the GRBs that take place outside 
the host galaxies do not produce significant afterglows, because
of low density of the ambient medium and therefore are not followed 
by afterglows.
This would make a strong selection effect against detecting 
GRBs happening outside of the hosts. At this conference we have
heard however, that the Beppo SAX  data are consistent with all
GRBs having X-ray afterglows \cite{Costa}. This means that all
GRBs have afterglows and that GRBs with no afterglows do not
exist, at least within the sample observable by Beppo SAX.

The reasoning presented above strongly argues against the 
compact object merger model of GRBs. However, we must remember
that afterglows have been detected only from the long bursts,
and we do not know if short bursts also produce afterglows
and what are the locations in relation to host galaxies of 
short bursts. Moreover, numerical
models of compact object coalescences \cite{kluzniak,ruffert}
agree with analytical estimates and show that the timescales
of these events can not be stretched beyond a fraction of a
second. Yet the long bursts have a median duration of
approximately $\approx 20 - 30\,$s.  

Thus we conclude that the compact object merger model appears
to be inconsistent with the observations of afterglows, and their
locations within the host galaxies. Long bursts are therefore most
probably not connected with compact object mergers. However,
it is quite likely that we will find that the short bursts 
are connected with mergers of compact objects.

{\bf Acknowledgments.} We acknowledge the support of the following
grants: KBN-2P03D01616, KBN-2P03D00415.


\begin{references}


\bibitem{bbz} Belczynski K., Bulik T. and Zbijewski W., {\it A\&A} accepted, 
        astro-ph/9911435 (1999).

\bibitem{bbzmn} Belczynski K., Bulik T. and Zbijewski W., {\it
MNRAS} {\bf 309}, 629 (1999).

\bibitem{bbrhere} Belczynski K., Bulik T. and Rudak B., 2000, this
volume.
	
\bibitem{cordeschernoff} Dewey R.J., Cordes J.M., {\it ApJ} {\bf 321}, 780
        (1987).
	
\bibitem{Costa} Costa E., 2000, this volume.
	
\bibitem{Fruchter} Fruchter A., 2000, this volume.

\bibitem{fryer98} Fryer C.L., Burrows A., Benz W., 
        {\it ApJ} {\bf 496}, 333 (1998). 
        
\bibitem{kluzniak} Lee W.H. and Klu{\'z}niak W., {\it Acta Astronomica} 
        {\bf 45}, 705 (1995).

\bibitem{ruffert} Ruffert M., Janka H.T., Takahashi K. and Schaefer G.,
        {\it A\&A} {\bf 319}, 122 (1997).


\end{references}
\end{document}